\documentclass[doublecol]{epl2}

\title{Modified Fluctuation-dissipation theorem for non-equilibrium steady-states
 and applications to molecular motors}
\shorttitle{Modified Fluctuation-dissipation theorem for non-equilibrium steady-states} 

\author{G. Verley\inst{1}  \and K. Mallick\inst{2} \and D. Lacoste\inst{1}}
\shortauthor{F. Author \etal}

\institute{
  \inst{1} Laboratoire de Physico-Chimie Th\'eorique - UMR CNRS Gulliver 7083,
ESPCI, 10 rue Vauquelin, F-75231 Paris, France\\
  \inst{2} Institut de Physique Th\'eorique, CEA Saclay, 91191 Gif, France\\
}

\abstract{We present a theoretical framework to understand a modified
fluctuation-dissipation theorem valid for systems close to non-equilibrium
steady-states and obeying markovian dynamics. We discuss the interpretation of
this result in terms of trajectory entropy excess.
The framework is illustrated on a simple pedagogical
example of a molecular motor. We also derive in this context
generalized Green-Kubo
relations similar to the ones obtained recently in
U. Seifert, Phys. Rev. Lett., \textbf{104}, 138101 (2010)
for more general networks of biomolecular states.}

\usepackage{graphicx}
\usepackage{amsmath}
\usepackage{amsfonts}
\usepackage{amssymb}

\newcommand{\bea}{\begin{eqnarray}}
\newcommand{\eea}{\end{eqnarray}}
\newcommand{\beq}{\begin{equation}}
\newcommand{\eeq}{\end{equation}}
\newcommand{\bit}{\begin{itemize}}
\newcommand{\eit}{\end{itemize}}

\newcommand{\ad}{\overrightarrow{\omega_a}}
\newcommand{\ag}{\overleftarrow{\omega_a}}
\newcommand{\bd}{\overrightarrow{\omega_b}}
\newcommand{\bg}{\overleftarrow{\omega_b}}

\begin{document}

							    \maketitle

\section{Introduction}
\label{Intro}

The application of linear response theory to systems in thermodynamic
equilibrium leads to the  fluctuation-dissipation theorem
(FDT) \cite{kubo66}, which states
that the response of an equilibrium system to small external
perturbations is determined by correlations {\it at} equilibrium.
 Suppose that a system at
 thermal  equilibrium and  governed by the   time-independent  Hamiltonian  $H_0$
 is subject to a time-dependent perturbation   $-\lambda(t) O$ from time   $t'$ on.
 Then the mean-value of a dynamic observable $A(t)$ at time $t >t'$ over all path trajectories,
 $\langle A(t) \rangle_{\rm path}$, satisfies at first
 order in $\lambda$:
\beq R_{eq} (t,t') =  \frac{\delta  \langle A(t) \rangle_{\rm path}}
 {\delta  \lambda(t')}
=  \beta \frac{d}{dt'} \langle  O(t')A(t)
\rangle_{eq},
\label{FDT_eq}
\eeq
where the correlation function in the r.h.s. is evaluated at equilibrium
 $\beta = 1/{k_B T}$ being  the inverse temperature.
This relation is a fundamental tool in statistical mechanics
since it allows to extract  linear response transport coefficients from an  equilibrium
 situation  \cite{lubensky,Rep-Vulpiani:2008}.
Beyond the equilibrium regime, the relation between  response and correlations
does not take a simple and universal form  as shown by   formal studies of such
 relations  for  stochastic processes
 \cite{Rep-hanggi:1982} or for glassy systems \cite{ritort-revueFDT}.
Experimentally, departures away from FDT in non-equilibrium systems
have been observed in a variety of systems such as granular matter,
sheared fluids and biological systems \cite{martin:2001}.

In the last decade,  new directions of
study on non-equilibrium systems have emerged.  For instance, it has been realized
that thermodynamic quantities like work \cite{Jarzynski:1997,ken:1998}
or entropy \cite{seifertPRL:2005} acquire a well defined meaning at
the level of  a single trajectory.  Various exact relations among the
statistical distributions of work or heat, called fluctuation
relations, have been derived. They typically hold very generally for a
large class of systems and arbitrarily far from equilibrium
\cite{GC:1995,Jarzynski:1997,crooks,lebowitz,hatano-sasa:2001}.
The entropy production has been
related in markovian systems to the difference between the forward and backward dynamical
randomness \cite{gaspard_Jstat:2004} or as the relative entropy
of the trajectory measures of the forward and backward dynamics \cite{lebowitz,chetrite:2008}.
For hamiltonian dynamics, similarly, the entropy production has been understood in terms of
the relative entropy between forward and backward probability distributions in phase space
\cite{kawai:2007}. A classification of the various possible decompositions of the entropy production and
of the corresponding fluctuation relations has been proposed \cite{Broek-esposito:2010}.  Within the linear
response regime and for slightly perturbed non-equilibrium steady
states (Ness), the fluctuation relations lead to a modified
fluctuation-dissipation theorem (MFDT)
\cite{seifert:2006,chetrite:2008,seifert-EPL:2010}, which has been
tested experimentally using colloidal particles in optical traps
\cite{chetrite-2009,blickle-2006}. A thermodynamic interpretation of MFDT using
the concept of entropy flow has been proposed in~\cite{maes:2009}.
Besides,  beyond the linear regime, the same
fluctuation relations can  be used to derive non-linear response
relations of higher order \cite{gaspard:2007}.

Let us consider a system initially in a
non-equilibrium steady state, characterized by a (set of) control parameters
denoted by  $\lambda$.
 For a given  value of
$\lambda$, we  assume  that there exists a steady state
with  stationary probability distribution
$P_{st}(c,\lambda) = \exp(-\phi(c,\lambda))$.
  A  time-dependent  perturbation
of the dynamics at time $t'$ around the fixed value $\lambda_{0}$  will be described by
 $\lambda(t')=\lambda_{0} + \delta \lambda(t')$.
 The response $R(t,t')  =\delta \langle A(c(t), \lambda_{0}) \rangle_{\rm path} / \delta
\lambda(t')$ of the  dynamic observable $A$ that depends on the microscopic
 configuration $c(t)$ at time $t>t'$  is given by the MFDT:
 \beq
 R(t,t') = - \frac{d}{dt'} \left \langle \left.
\frac{\partial \phi(c(t') , \lambda)}{\partial \lambda} \right|_{\lambda=\lambda_0} A( c(t),
\lambda_{0} ) \right \rangle,
\label{GFDT}
\eeq
 where $\langle ..\rangle$ denotes the average in the stationary state with the
 control parameter $\lambda_0$.
The relation (\ref{GFDT}) has been derived in the recent Ref.~\cite{prost-parrondo:2009}
for the particular observable  $A(c ,\lambda)=\partial \phi(c,\lambda)/\partial \lambda$, and
before that in Ref.~\cite{chetrite:2008} (relation 7.15) for the particular case of diffusion processes.
 We also  note   that  in  Eq.~\eqref{GFDT}, the function $\phi(c,\lambda)$
 plays the role of the  energy. For  thermal
 equilibrium, we have   $\phi(c,\lambda)=\beta
(H(c)-\lambda O(c)-F(\lambda))$,  where $F(\lambda)$ is the free
energy and   Eq.~(\ref{FDT_eq})  is
retrieved   [using the abbreviation $O(t')=O(c(t'))$].

Modified fluctuation-dissipation theorems
have appeared in various forms in the recent literature \cite{seifert-EPL:2010,maes:2009,Rep-Vulpiani:2008,chetrite:2008}.
In the first section of this paper, we present an elementary and self-contained
 derivation of such a result, which holds for any single-time
 observable $A(t)$ and for systems close
to non-equilibrium steady-states and obeying markovian dynamics.
In the second section, we discuss the interpretation of this relation in terms of trajectory entropy excess,
and finally we apply this framework to a simple model of molecular motor.

\section{Derivation of a modified fluctuation-dissipation theorem}
\label{Preuve}

We  consider a system which evolves according to a continuous time Markovian
dynamics. The transition rate  from a configuration $c$
to a configuration $c'$ is   denoted  by
$W_\lambda(c',c)$ to emphasize its dependence  on the  control  parameter $\lambda(t)$
 which can vary with time.
 For each  path trajectory,  we introduce, as in~\cite{hatano-sasa:2001},
  the  functional $Y(t)$  given by
 \beq Y(t)=\int_0^t \dot
\lambda(\tau) \frac{\partial \phi(c(\tau),\lambda(\tau))}{\partial
\lambda} d\tau.
\label{def Y(t)}
\eeq
 $Y(t)$  plays a role similar to the work in the Jarzynski relation
\cite{jarzynski-PRE:1997}.
The  joint probability
 \beq
P_t(c,Y)=\langle \delta (c-c(t))
\delta (Y-Y(t)) \rangle_{\rm path}
\eeq
for the system to be in configuration $c$ at
time $t$ with  $Y(t) = Y$  evolves according to
\beq \frac{\partial P_t(c,Y)}{\partial t}=
\sum_{c'} W_\lambda(c,c') P_t(c',Y) - \dot \lambda \frac{\partial
\phi(c,\lambda)}{\partial \lambda} \frac{\partial P_t(c,Y)}{\partial
Y}.
\eeq
The Laplace transform of $P_t(c,Y)$, given by
$\hat{P_t}(c,\gamma)=\int dY P_t(c,Y) e^{-\gamma Y}$,
 obeys the modified master equation:
\beq \frac{\partial \hat{P_t}}{\partial t}=\sum_{c'}
W_\lambda^\gamma(c,c') \hat{P_t}(c') =  W_\lambda^\gamma \cdot
\hat{P_t},
\label{dynamic evolution}
\eeq where $W_\lambda^\gamma$ is the matrix of elements \beq
W_\lambda^\gamma(c,c')=W_\lambda(c,c')-\dot \lambda \gamma
\frac{\partial \phi}{\partial \lambda} \delta_{c,c'}.
\label{Master eq Laplace}
\eeq
For a fixed value of $\lambda$ there exists a   stationary state $P_{st}$ such that
$W_\lambda \cdot P_{st} =0$.  Then, it can be checked directly that
the "accompanying" distribution (first defined in Ref.~\cite{Rep-hanggi:1982})
$P_{st}(c,\lambda(t))=e^{- \phi(c,\lambda(t))}$, solves Eq.~(\ref{dynamic evolution})  for $\gamma=1$. Note  that
this  "accompanying" distribution  $P_{st}(c,\lambda(t))$ is not stationary because
 it acquires a time dependence
through $\lambda(t)$. Therefore,  we have $\hat{P_t}(c,1) = e^{- \phi(c,\lambda(t))}$,
 or equivalently
\beq \langle \delta(c-c(t)) e^{-Y(t)}
\rangle_{\rm path}=e^{-\phi(c,\lambda(t))}.
\label{key relation}
\eeq
 We emphasize that the  l.h.s. depends on the full path history between
 time 0 and $t$, because $c(t)$ and $Y(t)$  do so, whereas the r.h.s.
 is a  function only  of the steady state probability corresponding to
 the value of  $\lambda$ at  the final  time $t$.
 This  relation  involves weighted averages with respect to the functional
 $e^{-Y(t)}$ and relates non-stationary expectation values to
 behavior in the stationary state.
 The use of   appropriately   weighted  distribution functions
 lies at  the core of  the various nonequilibrium  identities, as  emphasized
  in the  very first works of
  C. Jarzynski~\cite{Jarzynski:1997,jarzynski-PRE:1997}
 (see also \cite{lebowitz,jarzynski-2005,imparato:2005}).
 The relation~(\ref{key relation}) will also
 play   a key role in deriving the modified FDT.
 Multiplying this equation
by an arbitrary observable $A(c,\lambda)$ and summing over all  microscopic
 configurations $c$, we obtain a detailed version of the Hatano-Sasa
 identity~\cite{hatano-sasa:2001}
 \beq
\langle A(c(t),\lambda(t)) e^{-Y(t)} \rangle_{\rm path} = \langle A(\lambda(t))
\rangle_{\rm Ness}.
\label{variant of Sasa}
\eeq
where $\langle .. \rangle_{\rm Ness}$ denotes the average in the stationary
state  at time $t$  with  control parameter  $\lambda(t)$.
We now take the functional derivative of this relation with respect
 to $\lambda(t')$ with $t' < t$ by considering
  a small variation in the vicinity of  the stationary state
  $\lambda (t') =   \lambda_{0} + \delta \lambda(t')$ with
 $\delta \lambda(t') \ll 1$  and  $\delta \dot\lambda(t') \ll 1$.
 Then,   $Y(t)$  being  small, we can write
  at first order   $e^{-Y(t)} \simeq 1 - Y(t)$.
  Taking into account that the
  functional derivative of the r.h.s. of Equation~(\ref{variant of Sasa})
 with respect to $\lambda(t')$ vanishes for   $t' < t$,  we obtain
 \beq
  \frac {\delta \langle A(c(t),\lambda(t)) \rangle_{\rm path}}
  {\delta \lambda(t')}
  =
 \frac {\delta \langle  Y(t) \,   A(c(t),\lambda(t))  \rangle_{\rm path} }
 {\delta \lambda(t')}
 \label{F derivative of 9}
 \eeq
 The functional derivative of the r.h.s.
   in the vicinity of  $\lambda_{0}$   contains only one term instead of two
 because Y(t) vanishes when  $\lambda (t')$ takes the constant value   $\lambda_{0}$.
 Using
 \beq
\frac{\delta Y(t)}{\delta \lambda(t')} \Big|_{\lambda_{0}} =  -\frac{d}{dt'}
  \frac{\partial \phi(c(t'),\lambda_{0})}{\partial \lambda} \, ,
\eeq
 we obtain
 \beq
 R(t,t')  =  -\frac{d}{dt'} \langle \frac{\partial \phi(c(t'),\lambda_{0})}
 {\partial \lambda}  A(c(t),\lambda_{0})\,  \rangle_{\lambda(t)=\lambda_{0}} \,.
\eeq
 In the expectation value the control parameter is  now fixed at  $\lambda_{0}$
 and Eq.~(\ref{GFDT}) is proved.  Introducing
 the  observable ${\mathcal O}(c) = - \partial_\lambda P_{st}(c)/ P_{st}(c)$,  Eq.~\eqref{GFDT}
  can be  rewritten as
\beq
 R(t,t') = -\frac{d}{dt'} \left \langle   A(t)  {\mathcal O}(t') \right \rangle \, .
\label{rewriting1}
\eeq



{\it Remark:}  More general  versions of the FDT, valid
 for an  arbitrary observable   $F[c,\lambda]$,
that depends on  the whole  path (and not on the final configuration only)
 can be derived~\cite{chetrite:2008}
 by  comparing the weights
 of  direct and reverse path
trajectories   and using a local  detailed balance condition,
  in the spirit of  \cite{crooks}. The
  fundamental relation~(\ref{key relation}) has to be replaced
 by
\beq \langle
F[c,\lambda] e^{-Y(t)} \rangle_{\rm path} = \langle \tilde{F}[c,\lambda]
\rangle^r_{\rm path},
\label{crooks}
\eeq
 where the tilde and the index $r$
  denotes an average with respect to reverse paths.
 We emphasize, however, that in the derivation given above of
 the relation~(\ref{GFDT}) no symmetry  property under  time-reversal
 has  been used.


\section{Connection between MFDT and entropy production}

An important step towards an unification of the various formulations of FDT for non-equilibrium systems comes from the realization that the MFDT can be given by a thermodynamic interpretation
in terms of trajectory entropy excess \cite{maes:2009,seifert-EPL:2010,seifert:2010}.
Recently, a new decomposition of the entropy production
has been introduced in Refs.~\cite{Broek-esposito:2010,Esposito-mukamel:2007} in a particularly clear way.
This motivated us to revisit the
derivation of the MFDT of Refs.~\cite{seifert-EPL:2010,seifert:2010} with
this formalism. As expected, the decomposition of the entropy production leads
to an MFDT which is the sum of an equilibrium part and an additive correction.

We now focus on individual stochastic trajectories taken by the system. Between the time $t=0$ and $t=T$, these trajectories can be represented by the set of discrete values $\mathcal{C}=\{c_0,c_1....c_N \}$ and
jumping times $\tau_i$. The system stochastic entropy is defined as $s(t)=-\ln P_t(c(t))$, as a trajectory dependent quantity with $c(t)$ taking values in $\mathcal{C}$ \cite{seifertPRL:2005}.
Following \cite{Esposito-mukamel:2007}, we define the rate of change of the excess entropy
\beq
\dot{s}_{ex}(t)= \sum_{i=1}^N \delta (t-\tau_i) \ln \frac{P_{st}(c_i,\lambda_{\tau_i})}{P_{st}(c_{i-1},\lambda_{\tau_i})},
\label{s_ex}
\eeq
where $\tau_i$ represents the time where the system jumps from state $c_{i-1}$ to state $c_{i}$.
It follows that the integral of $\dot{s}_{ex}(t')$ from $t'=0$ to $t$, $\Delta s_{ex}(t)$ corresponds to
the excess heat defined in \cite{hatano-sasa:2001}, which satisfies $\Delta s_{ex}(t)= Y(t)- \Delta \phi(t)$ where $\Delta \phi(t) = \phi(c(t),\lambda(t)) - \phi(c(0),\lambda(0))$.

On a trajectory where $\lambda$ is fixed at $\lambda_0$,
\bea
\left. \frac{\partial \dot{s}_{ex}(t)}{\partial \lambda}  \right|_{\lambda=\lambda_0}
&=& \sum_{i=1}^N \delta (t-\tau_i) \frac{\partial} {\partial \lambda} \ln \frac{P_{st}(c_i,\lambda_0)}{P_{st}(c_{i-1},\lambda_0)}, \nonumber \\
 &=& \frac{d}{dt} \sum_n \delta_{c(t)n} \frac{\partial} {\partial \lambda} \ln P_{st}(n,\lambda_0), \nonumber \\
 &=& -\frac{d}{dt} \frac{\partial \phi(c(t),\lambda_0)} {\partial \lambda}.
 \label{echange}
\eea
After moving the time derivative in the r.h.s. of Eq.~\ref{GFDT} into the correlation function, we can then use Eq.~\ref{echange}
to obtain another formulation of MFDT:
\beq
R(t,t')= \left \langle \frac{\partial {\dot s}_{ex}(t')}{\partial \lambda} A(t) \right \rangle.
\label{new FDT}
\eeq

As shown in Refs.~\cite{Broek-esposito:2010,Esposito-mukamel:2007}, the excess entropy can be decomposed as $\dot{s}_{ex}=\dot{s}_r - \dot{s}_a$, where $s_r$ is the reservoir entropy
and $s_a$ the adiabatic entropy (also called house-keeping heat \cite{hatano-sasa:2001}). These quantities satisfy:
\bea
\dot{s}_{r}(t) &=& \sum_{i=1}^N \delta (t-\tau_i) \ln \frac{W_{\lambda_{\tau_i}}(c_i,c_{i-1})}{W_{\lambda_{\tau_i}}(c_{i-1},c_i)}, \nonumber \\
\dot{s}_{a}(t) &=& \sum_{i=1}^N \delta (t-\tau_i) \ln \frac{W_{\lambda_{\tau_i}}(c_i,c_{i-1}) P_{st}(c_{i-1},\lambda_{\tau_i})}{W_{\lambda_{\tau_i}}(c_{i-1},c_i) P_{st}(c_{i},\lambda_{\tau_i})}. \nonumber
\eea
In the stationary state (NESS) at $\lambda=\lambda_0$, it follows from Eq.~\ref{s_ex} that $\dot{s}_{ex}=-\dot{s}$, and thus Eq.~\ref{new FDT} agrees with Eq.~17 of Ref.~\cite{seifert-EPL:2010}. This also implies $\dot{s}_{na}=0$, and $\dot s_a=\dot s_{tot}$, and since $\dot s_r=s_{med}$, the splitting of the entropy excess which is used here is the same as that of Ref.~\cite{seifert-EPL:2010}.

We now proceed in deriving another form of MFDT with this framework. We assume that the system satisfies a generalized detailed balance condition
\beq
\frac{W_{\lambda}(c,c')}{W_{\lambda}(c',c)}=
\frac{W_{\lambda_0}(c,c')}{W_{\lambda_0}(c',c)} \exp \left( \delta \lambda d(c,c') \right),
\label{GDB}
\eeq
where $d(c,c')$ describes the variation of a dimensionless
physical quantity during a transition from state $c'$ to state $c$ such that $d(c,c')=-d(c',c)$ \cite{seifert:2010}. Using Eq.~\ref{GDB} and the definition of $\dot s_r$ one obtains
\beq
 \partial_\lambda \dot s_r (t') = \sum_{i=1}^N \delta(t'-\tau_i) d(c_i,c_{i-1})=j(t'),
\label{srp}
\eeq
where $j(t')$ corresponds to a physical current. Similarly, one can define $ \nu(t')= \partial_\lambda \dot s_a (t')$, in such a way that the response function takes the form
\beq
R(t-t') = \langle A(t) (j(t') - \nu(t')) \rangle. \label{repjnu}
\eeq
This form is analogous to the one first obtained for a particle obeying Langevin dynamics \cite{seifert:2006}, which has the property that  an equilibrium form of FDT can be restored in a locally moving frame \cite{chetrite:2008}. However, it is important to realize that the $\nu$ introduced above is different from the mean local velocity used in these references, although both quantities lead to the same correlation function \cite{seifert-EPL:2010}.

For practical applications of this result, more explicit expressions
of the currents $j(t')$ and $\nu(t')$ are needed.
For the part of the response function coming from the reservoir entropy (the equilibrium part), we can write
\beq
\begin{split}
&\langle A(t) j(t') \rangle  =  \langle A(t) \sum_{i=1}^N \delta(t'-\tau_i) d(c_i,c_{i-1}) \rangle,  \nonumber \\
& = \! \! \sum_{c,c',n} A_n P(n,t | c,t'^+) P(c,t'^+ | c',t'^-) P_{st}(c',\lambda_0) d(c,c'),
\end{split}
\eeq
which corresponds to a sum over trajectories which jump at time $t'$. We have denoted $P(n,t | c,t'^+)$
the conditional probability to be in state $n$ at time $t$ provided that the state $c$ was visited immediately after the jump at time $t'^+$. This quantity needs to be evaluated at $\lambda(t')=\lambda_0$.
Then, it follows that $ \langle A(t) j(t') \rangle $ is equal to
\beq
\begin{split}
\sum_{c,c',n} \! \! A_n \langle \delta_{c(t)n}  \delta_{c(t')c} \rangle \frac{W_{\lambda_0}(c,c')}{P_{st}(c,\lambda_0)} P_{st}(c',\lambda_0) d(c,c')\\
 =  \langle A(t) \sum_{c} \delta_{c(t')c} j(c,\lambda_0) \rangle,
 \end{split}
\eeq
where $j(c,\lambda_0)$ are components of the current $j(t')$ defined by
\beq
j(c,\lambda_0)=\sum_{c'} \frac{P_{st}(c',\lambda_0)}{P_{st}(c,\lambda_0)} W_{\lambda_0} (c,c') d(c,c'). \label{current}
\eeq

A similar calculation can be carried out for the part of the response function associated with the adiabatic entropy:
\beq
\langle A(t) \nu(t') \rangle = \langle A(t) \sum_{c} \delta_{c(t')c} \nu(c,\lambda_0) \rangle,
\eeq
where the components of the local current $\nu(t')$ are given by
\beq
\nu(c,\lambda_0)= \sum_{c'} \frac{J_{st}(c',c)}{P_{st}(c,\lambda_0)} \partial_\lambda \ln W_{\lambda_0}(c',c), \label{loc_v}
\eeq
and $J_{st}(c',c)$ denotes the probability current $P_{st}(c,\lambda_0) W_{\lambda_0}(c',c) - P_{st}(c',\lambda_0) W_{\lambda_0}(c,c')$. Note that Eq.~\ref{loc_v} agrees with the results given in Ref.~\cite{seifert:2010}.

\section{A discrete ratchet model}
\label{exemple}

We now apply the framework developed above to a discrete ratchet model of
a molecular motor.  Single molecular motors have been traditionally modeled
 either by continuous models such as the flashing ratchet model \cite{armand1}
 or by discrete models based on the master equation formalism \cite{kolomeisky-revue}. In previous
works, we have shown that the Gallavotti-Cohen symmetry is present both
in discrete models \cite{pre-FT,prl-FT} and in continuous ones
 \cite{flashing-ratchet} when all the relevant variables are taken into account.

\begin{figure}
\qquad \quad \includegraphics[scale=1]{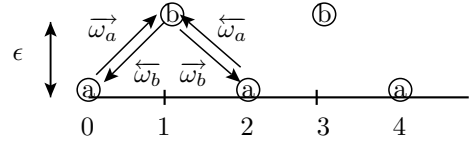}
\caption{A schematic representation of the motor on a linear lattice of sites $a$ (even) and $b$ (odd).
All possible transitions are displayed with their corresponding rates.}
\label{fig:sketch}
\end{figure}

In the discrete ratchet model, a single motor evolves on a
linear discrete lattice by hopping from one
site to neighboring sites, either consuming or producing ATP
molecules as shown in figure \eqref{fig:sketch}.
The position of the motor is denoted by $x = n d_0$, where
$2 d_0$ is the step size of the motor, and $y$ denotes the number of ATP molecules consumed.
Because of the periodicity of the filament, all the even
($a$) sites and all the odd ($b$) sites are equivalent.  Denoted by
$\overleftarrow{\omega}_{a}$ (and $\overrightarrow{\omega}_{a}$) are
the transition rates  for the motor  to jump from site $a$ to the
neighboring site $b$ to the left (to the right), respectively. A
similar definition holds for the site $b$ and we use the abbreviations
$\omega_i=\overleftarrow{\omega}_i + \overrightarrow{\omega}_i$ for
$i=a,b$, and $\Omega=\omega_a+\omega_b$.

The probability to find the motor in a given state, say $i=a,b$ is
$P_i(t)=\langle \delta_{ic(t)} \rangle_{\rm path}$, where $c(t)$ is the
configuration of the system at time $t$ in the space of configuration
${a,b}$. Similarly, the joint probability to be in state $i$ at time
$t$ and in state $j$ at time $t'$ is $P(i,t ; j,t')=\langle
\delta_{ic(t)} \delta_{jc(t')} \rangle_{\rm path}$. Both quantities can be
calculated analytically for this model {\it even for time dependent
rates}. We now assume that the rates depend on time only via an
arbitrary controlled parameter $\lambda(t)$, and we note that this
dependence can be non-linear.  As a result, the time dependance of an
arbitrary observable $A(c(t),\lambda)$ has the form \beq
A(c(t),\lambda)=A_a(\lambda) \delta_{ac(t)} + A_b(\lambda)
\delta_{bc(t)}.
\label{def A}
\eeq In particular, the function $\phi(c(t),\lambda)$ has this
form with $\phi_a (\lambda) =-\log
P_{st}(a,\lambda)=-\log [\omega_b(\lambda)/\Omega(\lambda)]$ and
$\phi_b(\lambda)=-\log P_{st}(b,\lambda)=-\log
[\omega_a(\lambda)/\Omega(\lambda)]$.

With the above equations, we can characterize the response of the
system to a perturbation of the rates of the form
$\omega_i(\lambda(t))=\omega_i(\lambda_{0}) + \delta \lambda(t)
\partial_\lambda \omega_i(\lambda_{0})$ for $i=a,b$.  We have
separately calculated both sides of Eq.~\eqref{GFDT}, and we found in
agreement with this equation the same quantity, which is the response
function associated with the observable $A(c,\lambda)$:
\begin{multline}
R(t,t')= \frac{\omega_a(\lambda_{0}) \partial_\lambda
\omega_b(\lambda_{0}) - \omega_b(\lambda_{0}) \partial_\lambda
\omega_a(\lambda_{0})}{\Omega(\lambda_{0})} \\ \left[
A_a(\lambda_{0})-A_b(\lambda_{0}) \right] \exp{\left[
-\Omega(\lambda_{0})(t-t') \right]},
\label{R_moteur}
\end{multline}
for $t>t'$.

\subsection{Decomposition of the response function}
\label{decomposition}

We now proceed in decomposing the above response function as a sum of two terms, which
correspond to the two parts of the entropy production discussed in the previous section.
In the following, we chose for the control
parameter either the normalized force applied on the motor, $f$, or
the normalized chemical potential difference associated with the ATP
hydrolysis reaction, $\Delta \mu$. These quantities are defined as $f=
F d_0/k_B T $ and $\Delta \mu=\Delta \tilde{\mu}/k_B T$, in terms of the
applied force $F$, and the chemical potential difference $\Delta
\tilde{\mu}$.  The sign convention for the force is such that it is
positive when it is in the motor motion direction.

In the case of a pure mechanical perturbation, $\lambda(t)=f(t)=f_{0}+\delta f(t)$.
The generalized detailed balance relations of Eq.~\ref{GDB} now takes the following
form:
\beq
\frac{\bd(f)}{\ag(f)}=\frac{\bd(0)}{\ag(0)}e^{f} , \;
\frac{\bg(f)}{\ad(f)}=\frac{\bg(0)}{\ad(0)}e^{-f},
\label{local DB discrete}
\eeq
with the correspondance $d(n \pm 1,n)= \pm 1$, valid for any position $n$.
These relations are obeyed by the following parametrization of the rates
\bea
\ad(f) &=& \omega e^{-\epsilon + \theta^+ f}, \;\;
\bd(f) =\omega' e^{ + (1-\theta^-) f}, \nonumber \\
\ag(f) &=& \omega' e^{-\epsilon - \theta^- f}, \;\;
\bg(f) =\omega e^{  -(1-\theta^+) f}, \label{Bilan_faible}\eea
where $\theta^+$ and $\theta^-$ are load distribution factors \cite{kolomeisky-revue}.

The motor velocity can be defined generally by $\langle v(t) \rangle =\sum_n n \partial P_n(t)/ \partial t$, with $P_n(t)$ the probability to find the motor on an integer position $n$ at time $t$. Since this velocity
is the current of the position variable, Eq.~\ref{current} can be used to define the components of this current:
\beq
v(a,f_0) = d_0 \frac{P_{st}(b,f_0)}{P_{st}(a,f_0)} \left[ \bd(f_{0}) -  \bg(f_{0}) \right], \nonumber \\
\eeq
and similarly for $v(b,f_0)$ by exchanging $a$ and $b$.
In a similar way, the components of the local current can be obtained from Eq.~\ref{loc_v}
\beq
\nu(a,f_0) = \frac{(\theta^+ + \theta^- )
\bar{v} }{2 P_{st}(a,f_{0})}, \quad \nu(b,f_0) = \frac{(2 - \theta^+ -
\theta^- ) \bar{v}  }{2 P_{st}(b,f_{0})}, \nonumber \eeq
with $\bar{v}=\langle v(t) \rangle=\langle \nu(t) \rangle$.
Now, as in Eq.~\eqref{repjnu}, we obtain the response function associated with an
observable $A$ :
\beq \frac{\delta \langle A (t)\rangle_{\rm path}}{\delta f(t')}=
\frac{1}{d_0}\langle A(t) ( v(t') - \nu(t') ) \rangle.
\eeq

For the case of a chemical perturbation in the concentrations of ATP,
or of ADP and P, the control parameter is $\lambda(t)=\Delta \mu(t)=\Delta \mu_{0}+
\delta \Delta \mu(t)$.  The transition rates for the motor to jump from a site $i$ to a neighboring site on the left or on the right with $l(=-1,0,1)$ ATP molecules consumed are $\omega_i^l = \overrightarrow{\omega_i}^l + \overleftarrow{\omega_i}^l$, with $i=a,b$. Local detailed balance conditions similar to Eq.~\eqref{local DB discrete}
 imply the following parametrization of the rates
\bea
\omega_a^1 &=& (\alpha + \alpha') e^{ -\epsilon + \sigma \Delta \mu} , \,\,\,\,\,\, \quad \omega_b^{-1} =(\alpha + \alpha') e^{ (\sigma -1) \Delta \mu}, \nonumber \\
\omega_a^0 &=& (\omega' + \omega) e^{-\epsilon}, \, \qquad \qquad \omega_b^0 = \omega + \omega ',
\eea
where $\sigma$ plays the same role as the $\theta^+$ and $\theta^-$ before.
Then, by a similar calculation, the response function can be written as
: \beq \frac{\delta \langle A (t)\rangle_{\rm path}}{\delta (\Delta \mu(t'))}=
\langle A(t) ( r(t')- \mathcal{R}(t')) \rangle, \eeq where $r(t')$
is the instantaneous ATP consumption rate and $\mathcal{R}(t')$ the
local ATP consumption rate, defined by their components
\begin{eqnarray*}
r(a,\Delta\mu_0) &=& - \omega_b^{-1}(\Delta \mu_0)P_{st}(b,\Delta\mu_0)/P_{st}(a,\Delta\mu_0),\\
r(b,\Delta\mu_0)&=& \omega_a^{1}(\Delta \mu_0)P_{st}(a,\Delta\mu_0)/P_{st}(b,\Delta\mu_0), \\
\mathcal{R}(a,\Delta\mu_0) &=& \sigma \bar r / P_{st}(a, \Delta \mu_0), \\
\mathcal{R}(b,\Delta\mu_0)&=& (1 -\sigma) \bar r / P_{st}(b, \Delta \mu_0).
\end{eqnarray*}

Now, we can also introduce more general rates, which depend on
both control parameters $f$ and $\Delta \mu$ \cite{pre-FT}.  The
method presented above in the particular cases where only a mechanical
degree of freedom or only a chemical degree of freedom is taken into
account, can be extended to more general situations where the state of
motor is described by both variables.  In this case, the same function
$\phi(c,\lambda)$ can be used, with the understanding that $c$
contains some dummy variables (the position variable $x(t)$ or the
chemical variable $y(t)$) in addition to the variables used to
describe the non-equilibrium steady state (namely $i=a,b$).  By
proceeding just as above, one obtains the response
functions in Eq.~\eqref{Green-Einstein1}, which take the form of modified Green-Kubo relations
\cite{seifert:2010,gaspard:2004}
\bea
 \langle v(t) \rangle_{\rm path} &-& \langle
v\rangle = \int_0^t \mathrm{d}t' \, \frac{\delta f(t')}{d_0}
\langle v(t)(v(t')-\nu(t')) \rangle  \nonumber \\
&+& \int_0^t \mathrm{d}t' \,
\delta \Delta \mu(t') \langle v(t)(r(t')-\mathcal{R}(t')) \rangle, \nonumber
 \\
\label{Green-Einstein1}
\langle r(t) \rangle_{\rm path} &-& \langle r \rangle = \int_0^t \mathrm{d}t' \, \frac{\delta f(t')}{d_0} \langle r(t)(v(t')-\nu(t')) \rangle  \nonumber \\
&+& \int_0^t \mathrm{d}t'
\, \delta \Delta \mu(t') \langle r(t)(r(t')-\mathcal{R}(t'))
\rangle.
\eea

\begin{figure}

\includegraphics[width=8cm,height=6cm]{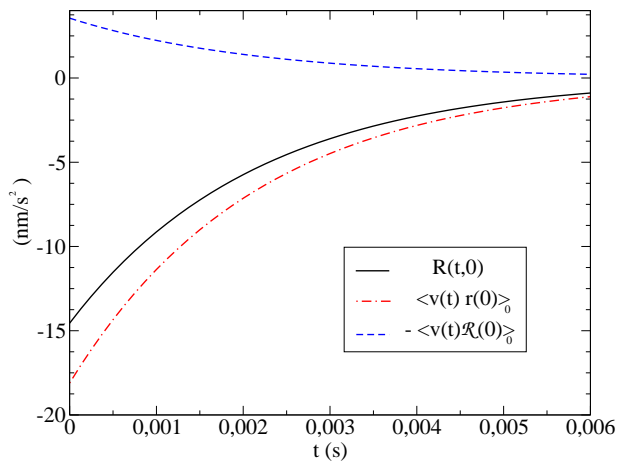}

\caption{Response function of the motor velocity $R(t,0)$, as function of time $t$ for a perturbation of the chemical potential $\Delta \mu$ applied at time 0 (solid line), correlation function of the velocity with the ATP consumption rate, $\langle v(t) r(0) \rangle$ (line-dotted curve), and correlation function of the velocity with the local ATP consumption rate $\langle v(t) \mathcal{R}(0) \rangle$ (dashed line). The initial condition corresponds to stalling for which $\bar v=0$. For these curves, we have used the following parameters: $\epsilon= 10.81 $, $ d_0=4 \, \mathrm{nm} $, $\omega =3.5 \, \mathrm{s}^{-1}$, $\omega'=108.15  \, \mathrm{s}^{-1}$, $\alpha = 0.57 \, \mathrm{s}^{-1}$, $\alpha'=1.3\times 10^{-6} \, \mathrm{s}^{-1}$, $\theta^{+}=0.705$, $\theta^{-}=1.375$, $\sigma=0.8$, $\Delta \mu = 11.8$ and $f=-3.82$ (stalling force). The sum of the two dashed lines gives the solid curve as imposed by Eq.~\eqref{Green-Einstein1}.}
\label{fig:Kubo}
\end{figure}

A few remarks about these equations are
in order: First, in the particular case of an equilibrium steady-state
(when $\nu=\mathcal{R}=0$), the Einstein and Onsager relations are
clearly recovered from these equations.  Secondly, near a non-equilibrium
steady state, these equations characterize the response of the motor
in the linear response regime, thus extending the results of
Ref.~\cite{pre-FT} to the case of time-dependent perturbations.  As
expected from the linearity of the problem, the response can be
decomposed as the sum of contributions corresponding to the cases of
pure mechanical and pure chemical perturbations.

Furthermore, we
note that the Einstein relation for the mechanical variable is
recovered only near stalling, just as in the case of time independent
perturbations \cite{prl-FT}. However, as pointed out in
Ref.~\cite{seifert:2010}, in more general networks of chemical
reactions, there are additional conditions besides the stalling
condition for the Einstein relation to hold.
In this model a mechanical perturbation applied to the motor
at stalling is thus unable to detect that the system is in a NESS. But, if
a perturbation in the more relevant chemical variable is considered, then
the NESS can be detected.
This point is illustrated in figure \eqref{fig:Kubo}, which
shows the deviation from the standard FDT (at equilibrium), deviation which
can be predicted from Eq.~\eqref{Green-Einstein1}.

\section{Conclusion}
\label{conclusion}

We have presented a general self-contained derivation of the modified FDT
for systems close to non-equilibrium steady-states and obeying markovian dynamics.
We believe that this derivation, which is related to many recent works on fluctuation relations,  is sufficiently general to lead to further developments.
We have also shown that the MFDT can be expressed as the correlation function of
a general observable with the trajectory entropy excess,
which leads to the decomposition of the MFDT into two terms.

We have applied this framework  to  a simple
model of molecular motor for which the steady-state probability distribution
is known analytically.
Finally, we have observed that the modified FDT relation
requires a knowledge of the relevant degrees of freedom in order
to be able to distinguish an equilibrium state from a non-equilibrium
steady state. In this choice of the relevant degrees of freedom, the
markovianity of the dynamics plays a central role, as it does for the
existence of a Gallavotti-Cohen symmetry.

\subsection{Acknowledgements}
We acknowledge stimulating discussions with U. Seifert, R. Chetrite,
T. Lubensky and R. Kawai.


\end{document}